\documentclass[11pt,showpacs,preprintnumbers,amsmath,amssymb,prd,nofootinbib,superscriptaddress]{revtex4-2}

\usepackage{dcolumn}% Align table columns on decimal point
\usepackage{bm}% bold math
\usepackage{ifpdf}
\usepackage{hyperref}
\usepackage{bm}
\usepackage{xcolor,color,graphicx,graphics,physics}
\usepackage[spanish,english]{babel}%, portuguese
\usepackage[latin1]{inputenc}
\usepackage[OT1]{fontenc}
\usepackage{latexsym,amssymb,amsmath,amsfonts, slashed}
\usepackage{makeidx}
\usepackage{epsfig,subfigure}
\usepackage{natbib}
\usepackage{epstopdf}
\usepackage{mathrsfs}
\usepackage{hyperref}%\usepackage[colorlinks=true,linkcolor=blue]{hyperref}
\hypersetup{colorlinks=true, linkcolor=blue, citecolor=blue, urlcolor=blue}
\usepackage{enumerate}

\usepackage{fixmath}

\everymath{\displaystyle}
\usepackage{graphicx}

\usepackage[T1]{fontenc}
\usepackage{amsmath}
\usepackage{amssymb}
\usepackage{graphicx}
\usepackage{xcolor}

\newcommand{\bea}{\begin{eqnarray}}
\newcommand{\eea}{\end{eqnarray}}

\newcommand{\orcid}[1]{\href{https://orcid.org/#1}{\includegraphics[width=10pt]{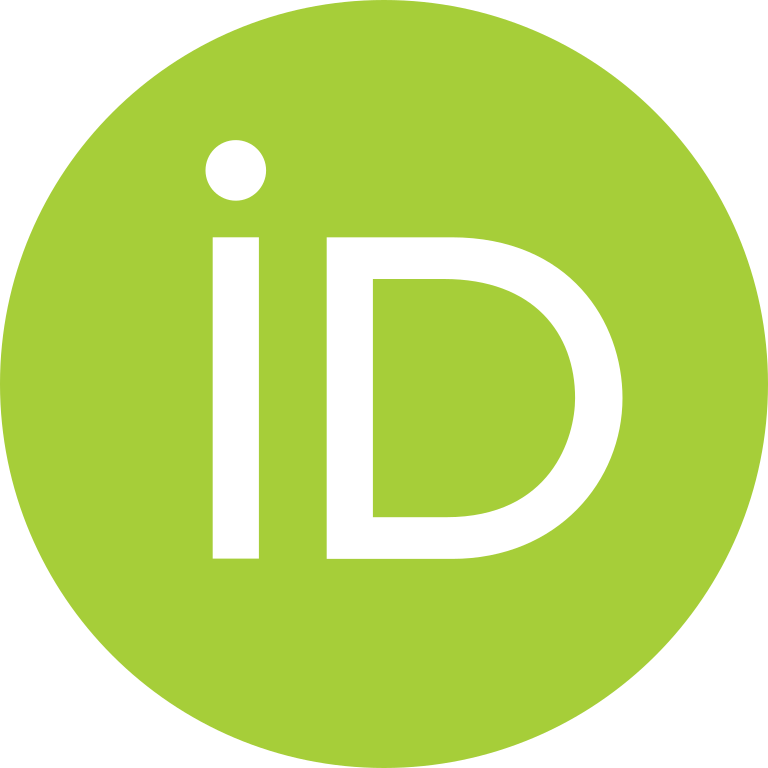}}}

\begin{document}

\title{Thermal Pair Production from Photon-Photon Collision:\\ Breit-Wheeler Process at Finite Temperature}

\author{D. S. Cabral  \orcid{0000-0002-7086-5582}}
\email{danielcabral@fisica.ufmt.br}
\affiliation{Instituto de F\'{\i}sica, Universidade Federal de Mato Grosso,\\
78060-900, Cuiab\'{a}, Mato Grosso, Brazil}

\author{A. F. Santos \orcid{0000-0002-2505-5273}}
\email{alesandroferreira@fisica.ufmt.br}
\affiliation{Instituto de F\'{\i}sica, Universidade Federal de Mato Grosso,\\
78060-900, Cuiab\'{a}, Mato Grosso, Brazil}

\author{R. Bufalo \orcid{0000-0003-1879-1560} }
\email{rodrigo.bufalo@ufla.br}
\affiliation{ Departamento de F\'isica, Universidade Federal de Lavras,\\
Caixa Postal 3037, 37200-900 Lavras, MG, Brazil}

\begin{abstract}

In this paper we examine the pair production through the Breit-Wheeler process $\gamma~\gamma \to e^+ e^-$ in a thermal background.
We compute the thermal contribution to the Breit-Wheeler differential cross section within the thermofield dynamics formalism.
We evaluate in details the cross section for this process, which possess a surprisingly simple expression valid for any temperature $\beta$, from which we discuss some physically relevant aspects. 
We also consider the high temperature regime of the cross section in order to have a better understanding about its thermal behavior.
\end{abstract}

\maketitle

\section{Introduction}

Although quantum electrodynamics (QED) can be regarded as one of the main triumphs in physics with unprecedented significance, by describing the atomic spectra and particle scattering, as well as being the cornerstone of the standard
model of particle physics, it is still an active area of research.
We can cite as one of the main ongoing interest the study of QED in intense electromagnetic fields, or simply strong-field QED \cite{Gonoskov:2021hwf,Fedotov:2022ely,DiPiazza:2011tq,Colgan:2022aax}, which had its beginning with the study of Schwinger-Sauter pair-production from an external field, which consists in the spontaneous creation of charged particle pairs by electromagnetic fields \cite{schwinger:1951,sauter:1931}.
On the experimental side of studying QED at extreme conditions, the continued development in laser technology and maximum focused intensities led to the growing impact of laser-plasma interactions
 in many facilities: SLAC \cite{Blumenfeld:2007ph}, the European XFEL \cite{Abramowicz:2021zja},  DESY \cite{Aschikhin:2015hya} and CERN \cite{AWAKE:2018gdq}.

In addition to the Schwinger's effect, one can also produce charged particle pairs from the collision of photons, for instance: Positron Production (from linear and nonlinear trident process) \cite{Burke:1997ew} and Breit-Wheeler process \cite{Breit:1934zz}.
The Breit-Wheeler pair production is a physical process in which a positron-electron pair is created from the collision of two photons, $\gamma~\gamma \to e^+ e^-$ (which is basically the inverse process of electron-positron annihilation), which has recently been experimentally observed \cite{STAR:2019wlg} \footnote{The detection of this Breit-Wheeler process, with quasi-real photons, occurred in the peripheral collisions of high Z nuclei at a conventional accelerator \cite{STAR:2019wlg}.}.
This process can be extended to the multiphoton (nonlinear) Breit-Wheeler process
\cite{ritus:1985,E144:1996enr}, when the collision of two photons is replaced by a high-energetic probe photon (a single $\gamma$ with several eV laser photons) decaying into pairs propagating through an electromagnetic field (for example, a laser pulse).
Both linear and nonlinear Breit-Wheeler processes have been extensively studied in many scenarios: 
extreme astrophysical environments, like those surrounding pulsars and gamma-ray bursts, but also in heavy-ion colliders, and high power laser facilities, among others  \cite{Bottcher:1997kf,Baur:1998ay,Klusek-Gawenda:2019ijn,Heinzl:2011ur,Blackburn:2014cig,DiPiazza:2016maj,Seipt:2016fyu,Titov:2018bgy,Ribeyre,Salgado:2021uua,Golub:2020kkc,Golub:2021nhj,Adamo:2021jxz,Kettle:2021ipe,Zhao:2021ynd,Blackburn:2021cuq,Brandenburg:2022tna,Turner:2022hch,MacLeod:2022qid}.

One interesting aspect to be analyzed, on theoretical grounds, in the production of charged pairs is  the role played by thermal effects and possible implications.
Although the Schwinger effect at finite temperature has been widely studied \cite{King:2012kd,medina:2017,gould:2017,brown:2018,gould:2018,korwar:2018,torgrimsson:2018,gould:2019,wang:2019}, as well as the Breit-Wheeler process \cite{torgrimsson:2018,gould:2019} (see ref.~\cite{torgrimsson:2018} for a detailed list of works) 
\footnote{These should not be thought of as a competing thermal processes, actually, applying the Optical Theorem, one can see that a unitarity cut of the dominant Feynman diagram for thermal Schwinger pair production gives the Breit-Wheeler process  \cite{gould:2019}.},
pair production in an electric field at finite temperature (with thermal fermions and/or thermal photons) has a long history with many conflicting papers
\cite{Fedotov:2022ely}.
Realizing that the thermal pair production is a somewhat controversial topic, in which many papers (results) disagree with each other, we approach the  Breit-Wheeler process in terms of the thermofield dynamics, with the expectation to shed some new light and clarify (some of) the controversies.

The close relation of the technical aspects of the  thermofield dynamics \cite{takahashi:1996,umezawa:1982,umezawa:1995,khanna:2009} with the case of $T = 0$ field theory, considering the analysis and calculations of scattering amplitudes and decay processes, makes this formalism very interesting to
describe finite temperature field theory, presenting some practical advantages in regard to Matsubara's imaginary time formalism and Schwinger-Keldysh closed time path method \cite{matsubara:1955,schwinger:1961,kapusta:2011}.
The main feature of the thermofield dynamics method is that the Fock space is doubled, in which a new set of operators is introduced, designed tilde operators, acting on the second Fock space, tilde space.
This can be understood as the following: 
physically speaking, the second Fock space is interpreted as a heat
bath that ensures the dynamical system to stay in equilibrium.
These two Fock spaces are connected by means of a Bogoliubov transformation, which 
consists of a rotation between the two spaces, original and tilde.

Since finite temperature effects have profound implications in the study of high energy physics, we examine the pair production of Breit-Wheeler process in the presence of a heat bath within the thermofield dynamics.
We start Sec. \ref{sec2} by reviewing the main aspects of the  Breit-Wheeler process and establishing the framework of thermofield dynamics to compute the respective amplitude related with the differential cross section.
In Sec. \ref{sec3} we evaluate explicitly the squared transition amplitude of the  Breit-Wheeler process in the center of mass frame, where the kinematics of the process takes a simple form.
Moreover, since the angular dependence is such that we can obtain the respective cross section valid for any value of $\beta$, from which we discuss some physically relevant aspects, including comments about the presence of off-shell contributions.
In Sec. \ref{sec4} we summarize the results, and present our final remarks.

\section{Breit-Wheeler Amplitude at finite temperature}
\label{sec2}

\begin{figure}[h]
    \centering
\includegraphics[scale=0.5]{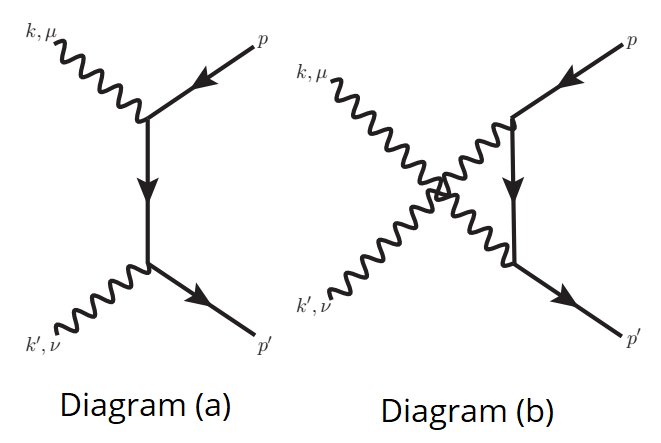}
\caption{The Feynman diagrams contributing to the pair creation in the BW process.}
\label{fig1}
\end{figure}

The main purpose of this section is to calculate the differential cross section for the Breit-Wheeler (BW) process at finite temperature within the thermofield dynamics (TFD) \cite{takahashi:1996,umezawa:1982,umezawa:1995,khanna:2009}.  The BW process consists of creating an electron-positron pair through the collision of two photons, i.e. $\gamma+\gamma=e^++e^-$.
The Feynman diagrams that describe this process are depicted in Figure \ref{fig1}. 

Since we wish to compute the thermal cross section of the BW process, our starting point is 
\begin{equation} \label{eq1}
    \frac{d\sigma}{d\Omega}=\frac{1}{64\pi^2(4\omega^2)}\langle|\widehat{\mathcal{M}}_{fi}|^2\rangle
\end{equation}
in which the temperature effects are solely contained in the transition amplitude $\mathcal{M} (\beta)$.
This thermal amplitude is formally defined as
\begin{equation}
\mathcal{\mathcal{M}}\left(\beta\right)=\left\langle f,\beta\left|\hat{S}\right|i,\beta\right\rangle 
\end{equation}
where the $\hat{S}$-matrix is given 
\begin{equation} \label{smatrix}
\hat{S}=\sum_{n=0}^{\infty}\frac{\left(-i\right)^{n}}{n!}\int dx_{1}dx_{2}\cdots dx_{n}\tau\left[\hat{\mathcal{L}}_{i}\left(x_{1}\right)\hat{\mathcal{L}}_{\rm int}\left(x_{2}\right)\cdots\hat{\mathcal{L}}_{\rm int}\left(x_{n}\right)\right],
\end{equation}
with $\tau$ being the time ordering operator and
\begin{equation}
\hat{\mathcal{L}}_{\rm int}\left(x\right) = \mathcal{L}_{\rm int}\left(x_{n}\right) - \tilde{\mathcal{L}}_{\rm int}\left(x_{n}\right)
\end{equation}
describes the Lagrangian interaction part in the doubled notation of the TFD formalism.
In this context, the final and initial thermal states for the process $\gamma (k)+\gamma (k')=e^+(p')+e^-(p)$ are given as
\begin{align} \label{initialstate}
\ket{i}&=\ket{\gamma,\gamma^\prime}=d^\dagger_\beta(k)d^\dagger_\beta(k^\prime)\ket{0(\beta)},\\  
  \ket{f}&=\ket{e^{+},e^{-}}=b^\dagger_\beta(p)a^\dagger_\beta(p^\prime)\ket{0(\beta)},\label{finalstate}
\end{align}
with $d^\dagger_\beta$, $b^\dagger_\beta$ and $a^\dagger_\beta$ being the thermal creation operators for the photons, electrons and positrons, respectively.

In order to calculate the thermal transition amplitude for the BW process, the second-order term of the scattering matrix \eqref{smatrix} is considered, 
\begin{equation}
    \widehat{S}^{(2)}=\int d^4xd^4y\bra{f} \tau \left[\hat{\mathcal{L}}_{\rm int}(x)\hat{\mathcal{L}}_{\rm int}(y)\right]\ket{i},
\end{equation}
and the thermal transition amplitude for the BW process is defined as
\begin{align}
\widehat{\mathcal{M}}_{fi}=\bra{f,\beta}\widehat{S}^{(2)}\ket{i,\beta}=&-\frac{e^2}{2}\int d^4xd^4y\left\{\bra{f}\tau \left[ \mathcal{L}_{\rm int}(x)\mathcal{L}_{\rm int}(y)\right]\ket{i}+  \bra{f}\tau \left[ \widetilde{\mathcal{L}}_{\rm int}(x)\widetilde{\mathcal{L}}_{\rm int}(y)\right]\ket{i}\right. \cr 
-& \left. \bra{f}\tau \left[\mathcal{L}_{\rm int}(x)\widetilde{\mathcal{L}}_{\rm int}(y)\right]\ket{i}-  \bra{f}\tau \left[ \widetilde{\mathcal{L}}_{\rm int}(x)\mathcal{L}_{\rm int}(y)\right]\ket{i}\right\}.\label{eq01}
\end{align}

In the quantum electrodynamics, the interaction part of the Lagrangian is given as
\begin{align}
\mathcal{L}_{\rm int}(x) &=e\Bar{\psi}(x)\gamma^\mu\psi(x)A_\mu(x), \\
\tilde{\mathcal{L}}_{\rm int}(x) &=e\tilde{\Bar{\psi}}(x)\gamma^\mu\tilde{\psi}(x)\tilde{A}_\mu(x),
\end{align}
where $e$ is the electron charge, $\gamma^\mu$ are the Dirac matrices, $\psi(x)$ is the fermion field and $A_\mu(x)$ is the photon field, while the tilde fields are the TFD dual (conjugated) fields.

In order to evaluate explicitly the matrix elements in \eqref{eq01}, we should also consider the solution for the fermionic field (and its tilde dual)
\begin{equation}
    \psi(x)=\int d^3pN_p\sum_s\left[b(p,s)u(p,s)e^{-ipx}+a^{\dagger}(p,s)\bar{u}(p,s)e^{ipx}\right],
\end{equation}
where $N_p$ is the normalization constant and $px=\xi t-\vec{p}\cdot\vec{x}$, with the dispersion relation $\xi= \sqrt{\vec{p}^2+m^2}$ and metric signature $\eta_{\mu\nu}= {\rm diag} (+1,-1,-1,-1)$.
In the TFD formalism it is worth to mention the (tilde) conjugation operations for the fermions $\tilde{u}(p,s)=u^{\dagger}(p,s)$.
Moreover, for the photon field
\begin{equation}
A_\mu(x)=\int d^3kM_k\sum_\lambda \left[d(k,\lambda)\epsilon_\mu(k,\lambda)e^{-ikx}+ d^{\dagger}(k,\lambda)\epsilon^{\dagger}_\mu(k,\lambda)e^{ikx}\right]
\end{equation}
with $\epsilon_\mu(k,\lambda)$ being the polarization vector of the physical polarization states.
With these elements we are now in a position to compute the amplitude \eqref{eq01}. 
Let us start with the first term
\begin{align}
\bra{f}\tau \left[\mathcal{L}_{\rm int}(x)\mathcal{L}_{\rm int}(y)\right]\ket{i}=~&2\Bar{u}(p^\prime,s^\prime)\gamma^\mu S(x-y)\gamma^\nu v(p,s)\varepsilon_\mu(k,\lambda)\varepsilon_\nu(k^\prime,\lambda^\prime)\nonumber\\
&\times\left[e^{i(p-k)x+i(p^\prime-k^\prime)y}+e^{i(p-k^\prime)x+i(p^\prime-k)y}\right] \nonumber
\\&\times\bra{f}\tau\left[a^\dagger(p^\prime,s^\prime)b^\dagger(p,s)d(k,\lambda)d(k^\prime,\lambda)\right]\ket{i}.\label{eq12}
\end{align}
One should pay attention to the element matrix present in \eqref{eq12} because it involves both $T=0$ and thermal creation (annihilation) operators (recall that the in and out states are thermal states, see eqs.~\eqref{initialstate} and \eqref{finalstate}).
In order to evaluate this element matrix we should use the relation among these two sets of operators, which formally is given in terms of the following Bogoliubov transformation \cite{khanna:2009}
\begin{equation}
a(p)=Ua_\beta(p)+V\widetilde{a}^\dagger_\beta(p);\quad\quad\quad a^\dagger(p)=Ua^\dagger_\beta(p)+V\widetilde{a}_\beta(p),
\end{equation}
for the case of fermions (analogous relations hold for $d(p,s)$) and
\begin{equation}
d(p)=U^\prime d_\beta(p)+V^\prime\widetilde{d}^\dagger_\beta(p);\quad\quad\quad d^\dagger(p)=U^\prime d^\dagger_\beta(p)+V^\prime\widetilde{d}_\beta(p)
\end{equation}
for bosons.
The set of functions $ (U, V)$ and $ (U',V')$ are related with the Fermi-Dirac $n_F (E)$ and Bose-Einstein $n_B (E)$ distributions, respectively, defined as the following
\begin{align}
U&=\frac{1}{\sqrt{1+e^{-\beta E}}}=\sqrt{e^{\beta E}n_{F}\left(E\right)},\quad V=\frac{1}{\sqrt{1+e^{\beta E}}}=\sqrt{n_{F}\left(E\right)}, \label{dist1}\\
U^{\prime}&=\frac{1}{\sqrt{1-e^{-\beta E}}}=\sqrt{e^{\beta E}n_{B}\left(E\right)},\quad V^{\prime}=\frac{e^{-\beta E/2}}{\sqrt{1-e^{-\beta E}}}=\sqrt{n_{B}\left(E\right)}.\label{dist2}
\end{align}
Thus, with help of these informations we cast the element matrix of Eq.~\eqref{eq12} as
\begin{align}
\bra{f}\tau\left[a^\dagger(p^\prime,s^\prime)b^\dagger(p,s)d(k,\lambda)d(k^\prime,\lambda)\right]\ket{i}=U^2(U^\prime)^2\bra{f}\tau\left[ a_\beta^\dagger(p^\prime,s^\prime)b_\beta^\dagger(p,s)d_\beta(k,\lambda)d_\beta(k^\prime,\lambda) \right]\ket{i},
\end{align}
in which all operators are now thermal and this element can be evaluated.

From this analysis we can compute the remaining matrix elements of Eq.~\eqref{eq01}, i.e. those proportional to $\left\langle\widetilde{\mathcal{L}}\widetilde{\mathcal{L}} \right\rangle$, $\left\langle\widetilde{\mathcal{L}}{\mathcal{L}}\right\rangle$ and $\left\langle{\mathcal{L}}\widetilde{\mathcal{L}} \right\rangle$, and it follows that they contribute with the following thermal functions $V^2(V^\prime)^2$, $-VU(V^\prime)(U^\prime)$ and $-UV(U^\prime)(V^\prime)$, respectively.
Hence, taking all these results into consideration, the transition amplitude \eqref{eq01} becomes in momentum space
\begin{align} \label{eq77}
\widehat{\mathcal{M}}_{fi}=&-ie^{2}F(\beta)\left(2\pi\right)^{4}\Big[\Bar{u}(p^{\prime},s^{\prime})\gamma^{\mu}S_{\beta}\left(k-p\right)\gamma^{\nu}v(p,s)\varepsilon_{\mu}(k,\lambda)\varepsilon_{\nu}(k^{\prime},\lambda^{\prime}) \cr
&+\Bar{u}(p^{\prime},s^{\prime})\gamma^{\mu}S_{\beta}\left(k'-p\right)\gamma^{\nu}v(p,s)\varepsilon_{\mu}(k,\lambda)\varepsilon_{\nu}(k^{\prime},\lambda^{\prime})\Big] \cr
&\times \Bar{u}(p^\prime,s^\prime)\gamma^\mu S_\beta(q) \gamma^\nu v(p,s)\varepsilon_\mu(k,\lambda)\varepsilon_\nu(k^\prime,\lambda^\prime)  \cr
&\times\bra{f}\tau\left[ a_\beta^\dagger(p^\prime,s^\prime)b_\beta^\dagger(p,s)d_\beta(k,\lambda)d_\beta(k^\prime,\lambda)\right]\ket{i},
\end{align}
where thermal distributions are contained in the factor 
\begin{equation}
    F(\beta)=(UU^\prime+VV^\prime)^2
\end{equation}
which can be written with help of \eqref{dist1} and \eqref{dist2} in terms of the thermal distributions
\begin{equation} \label{eq151}
    F(\beta)=\frac{1}{2} \left[\tanh \left(\frac{\beta  \omega }{2}\right)+2 \sqrt{\frac{1}{e^{\beta  \omega }-1}}
   \sqrt{\frac{1}{\cosh (\beta  \omega )+1}} \sqrt{\coth \left(\frac{\beta  \omega }{2}\right)+1}+\coth \left(\frac{\beta  \omega }{2}\right)\right].
\end{equation}
Moreover, we have that $S_{\beta}(q)$ is the fermion propagator at finite temperature defined as
\begin{equation}
S_\beta(q)=S^{(0)}(q)+S^{(\beta)}(q),
\end{equation}
in which $S^{(0)}(q) = \frac{\slashed{q}+m}{q^2-m^2}$ is the usual $T=0$ propagator, while $S^{(\beta)}(q)$ is the finite temperature defined as \cite{khanna:2009}
\begin{align}
 S^{(\beta)}(q)=\frac{\pi i}{(e^{\beta q^0}+1)}\left[\frac{(\gamma^0\xi-\gamma^{i}q_i+m)}{2\xi}\Delta_1\delta(q^0-\xi)+\frac{(\gamma^0\xi+\gamma^{i}q_i-m)}{2\xi}\Delta_2\delta(q^0+\xi)\right],
\end{align}
with the definitions
\begin{equation}
    \Delta_1=\begin{pmatrix}
    1 & e^{\beta q_0/2}\\e^{\beta q_0/2}&-1\end{pmatrix};\quad\quad \Delta_2=\begin{pmatrix}
    -1 & e^{\beta q_0/2}\\e^{\beta q_0/2}&1\end{pmatrix}.
\end{equation}

Finally, taking these results into account, we are able to compute the squared modulus of the transition amplitude
\begin{align} \label{eq78}
    \langle|\widehat{\mathcal{M}}_{fi}|^2\rangle=F^2(\beta)\left\{\langle|\widehat{\mathcal{M}}_{a}|^2\rangle_\beta+\langle|\widehat{\mathcal{M}}_{b}|^2\rangle_\beta+\left\langle2\Re\left[\widehat{\mathcal{M}}_{a}^\dagger\widehat{\mathcal{M}}_{b}\right]\right\rangle_\beta\right\}
\end{align}
with
\begin{align}
\langle|\widehat{\mathcal{M}}_{a}|^2\rangle_\beta =&\frac{e^4}{4}\eta_{\mu\rho}\eta_{\nu\alpha}\Tr{(\slashed{p}^{\prime}+m)\gamma^\nu S_\beta(k-p)\gamma^\mu(\slashed{p}-m)\gamma^\rho S_\beta^\dagger(k-p)\gamma^\alpha};\label{21}\\
\langle|\widehat{\mathcal{M}}_{b}|^2\rangle_\beta =&\frac{e^4}{4}\eta_{\nu\rho}\eta_{\mu\alpha}\Tr{(\slashed{p}^{\prime}+m)\gamma^\mu S_\beta(k^\prime-p)\gamma^\nu(\slashed{p}-m)\gamma^\rho S_\beta^\dagger(k^\prime-p)\gamma^\alpha};\label{22}\\
\left\langle2\Re\left[\widehat{\mathcal{M}}_{a}^\dagger\widehat{\mathcal{M}}_{b}\right]\right\rangle_\beta=&\frac{e^4}{2}\eta_{\mu\rho}\eta_{\nu\alpha}\Re{\Tr\left[(\slashed{p}-m)\gamma^\rho S_\beta(k-p)\gamma^\alpha(\slashed{p}^\prime+m)\gamma^\mu S_\beta^\dagger(k^\prime-p)\gamma^\nu\right]}.\label{23}
\end{align}
where we have averaged over the polarization and spin states, with use of the completeness relations, 
\begin{equation}
\sum_{s}u(p,s)\bar{u}(p,s)={\slashed{p}+m},\quad\quad\sum_{s}v(p,s)\bar{v}(p,s)={\slashed{p}-m}
\end{equation}
and 
\begin{equation}
\sum_{\lambda}\epsilon_\mu(k,\lambda)\epsilon^{\dagger}_\rho(k,\lambda)=\eta_{\mu\rho}.
\end{equation}
Our next step to compute the cross section for the Breit-Wheeler process is to compute explicitly \eqref{eq78}.

\section{Cross section at finite temperature}
\label{sec3}

In order to evaluate explicitly the squared transition amplitude \eqref{eq78} we consider the center of mass frame, so that the kinematical variables are defined as
\begin{align}
p =&\,(\omega,A\sin{\theta},0,A\cos{\theta}),\qquad  p^\prime=(\omega,-A\sin{\theta},0,-A\cos{\theta}) \\
k=&\, (\omega,0,0,\omega),\qquad k^{\prime}=(\omega,0,0,-\omega),
\end{align}
where $A=\sqrt{\omega^2-m^2}$, in this case the Mandelstam variables become
\begin{equation}
    s=2m^2-t-u, \qquad t=m^2-2\omega(\omega+A\cos{\theta}), \qquad u=m^2-2\omega(\omega-A\cos{\theta}).
\end{equation}

Within this framework we can now evaluate the trace in Eqs.~\eqref{21}, \eqref{22} and \eqref{23}, so that the transition amplitude \eqref{eq78} is cast as
\begin{equation}
\langle|\widehat{\mathcal{M}}_{fi}|^2\rangle=e^4F^2(\beta)\left[\mathcal{M}^{(0)}+\frac{1}{(1+e^{2\beta\omega})^2} \mathcal{M}^{(\beta)}\right]
\end{equation}
in which the thermal factor $F(\beta)$ is given by \eqref{eq151}, and the $T=0$ part reads
\begin{equation}
    \mathcal{M}^{(0)}=\frac{-8 m^4-4 \cos ^4(\theta ) \left(m^2-\omega ^2\right)^2+8 m^2 \cos ^2(\theta ) \left(m^2-\omega ^2\right)+8 m^2 \omega ^2+4 \omega ^4}{\left[\cos ^2(\theta
   ) \left(m^2-\omega ^2\right)+\omega ^2\right]^2}\label{30}
\end{equation}
while the thermal amplitude has the following expression
\begin{align}
\mathcal{M}^{(\beta)}=&-4\pi\frac{\Gamma_1+3\xi_t^2m^2}{\xi_t^2 }\left[\delta^2(2\omega+\xi_t)+\delta^2(2\omega-\xi_u)\right]\nonumber\\
&-4\pi\frac{\Gamma_1+3\xi_u^2m^2}{\xi_u^2 }\left[\delta^2(2\omega+\xi_u)+\delta^2(2\omega-\xi_u)\right]\nonumber\\
&+8\pi^2\frac{\left(3\xi_t^2m^2-\Gamma_1\right)}{\xi_t^2 }\delta(2\omega+\xi_t)\delta(2\omega-\xi_t)+8\pi^2\frac{\left(3\xi_u^2m^2-\Gamma_1\right)}{\xi_u^2 }\delta(2\omega+\xi_u)\delta(2\omega-\xi_u)\nonumber\\
&+8\pi^2\frac{\Gamma_2+(m^2-4\omega^2)\xi_t\xi_u}{\xi_t\xi_u }\left[\delta(2\omega+\xi_t)\delta(2\omega+\xi_u)+\delta(2\omega-\xi_t)\delta(2\omega-\xi_u)\right]\nonumber\\
&+8\pi^2\frac{\Gamma_2-(m^2-4\omega^2)\xi_t\xi_u}{\xi_t\xi_u }\left[\delta(2\omega+\xi_t)\delta(2\omega-\xi_u)+\delta(2\omega-\xi_t)\delta(2\omega+\xi_u)\right],\label{31}
\end{align}
where we have defined, by simplicity of notation, the following quantities
\begin{align}
    \Gamma_1&=18 m^4-19 \omega ^2 m^2-18 \omega  m^2\sqrt{\omega ^2-m^2} \cos (\theta )-\omega ^2m^2 \cos (2 \theta ) -\omega
   ^4+\omega ^4 \cos (2 \theta ),\\
   \Gamma_2 &=6 m^4-8 \omega ^2 m^2+8 \omega ^4+2 \omega  \sqrt{\omega ^2-m^2} \left(4 \omega ^2-3 m^2\right) \cos(\theta ), \\
   \xi_{\text{t}}&=\sqrt{2 \omega  \cos (\theta ) \sqrt{\omega ^2-m^2}+2 m^2-2 \omega ^2},\\
   \xi_{\text{u}}&=\sqrt{-2 \omega  \cos (\theta ) \sqrt{\omega ^2-m^2}+2 m^2-2 \omega ^2}.
\end{align}

Hence, in terms of the definition \eqref{eq1} the differential cross section for the Breit-Wheeler process at finite temperature is given as
\begin{equation} \label{eq181}
    \frac{d\sigma}{d\Omega}=\frac{1}{64\pi^2(4\omega^2)}\langle|\widehat{\mathcal{M}}_{fi}|^2\rangle=\frac{e^4F^2(\beta)}{256\pi ^2\omega^2}\left[\left.\frac{d\sigma}{d\Omega}\right|_{(0)}+\frac{1}{(1+e^{2\beta\omega})^2} \left.\frac{d\sigma}{d\Omega}\right|_{(\beta)}\right]
\end{equation}
where the contribution at $T=0$ is given by
\begin{equation}
\left.\frac{d\sigma}{d\Omega}\right|_{(0)}= \mathcal{M}^{(0)},
\end{equation}
while the finite temperature contribution is
\begin{equation}
\left.\frac{d\sigma}{d\Omega}\right|_{(\beta)}= \mathcal{M}^{(\beta)}
\end{equation}
where the zero and finite temperature amplitudes $\mathcal{M}^{(0)}$ and $\mathcal{M}^{(\beta)}$ are given by Eqs. \eqref{30} and \eqref{31}, respectively.

Some comments are in order about the expression \eqref{eq181}:
it is interesting to observe that at the limit when the temperature goes to zero ($\beta\rightarrow \infty$) it implies that  $\frac{1}{(1+e^{2\beta\omega})^2} \to0$ and $F(\beta)\to1$, then, differential cross section \eqref{eq181} is purely the known result at zero temperature \cite{Bottcher:1997kf,torgrimsson:2018,Ribeyre}.

Since all angular dependence of the differential cross section \eqref{eq181} is present in the  amplitudes squared $\mathcal{M}^{(0)}$ and $\mathcal{M}^{(\beta)}$, one can evaluate the angular integrations and obtain the total cross section for any value of $\beta$
\begin{equation} \label{eq_total_CS}
\sigma_{\rm BW}	 	=\frac{e^{4}}{32\pi\omega^{2}}F^{2}\left(\beta\right)\left[\mathbb{I}_{1}+\frac{1}{\left(1+e^{2\beta\omega}\right)^{2}}\mathbb{I}_{2}\right]
\end{equation}
in which it was defined 
\begin{align}
\mathbb{I}_{1}  =-\frac{2}{\chi}-2+\frac{3\chi+\frac{1}{\chi}-2}{\sqrt{\chi\left(\chi-1\right)}}\ln\left(\frac{1+\sqrt{1-\frac{1}{\chi}}}{1-\sqrt{1-\frac{1}{\chi}}}\right)
\end{align}
and
\begin{align}
\mathbb{I}_{2}	= &\frac{ \pi^{2}m}{\sqrt{\chi\left(\chi-1\right)}}\left(-10+21\chi-8\chi^{2}\right)\delta\left(\sqrt{\chi}\right) \cr
&	+\frac{4\pi^{2}}{\sqrt{\chi\left(\chi-1\right)}}\left(9+20\chi^{2}-24\chi\right)\delta\left(3\chi-1\right)
\end{align}
in which we have introduced the parameter  $\chi=\frac{\omega^{2}}{m^{2}}$ by means of simplicity of notation. One can observe that the term $\mathbb{I}_{1}$ corresponds to the usual $T=0$ contribution, while $\mathbb{I}_{2}$ corresponds to the purely thermal part.

It is important to observe that the delta functions present in $\mathbb{I}_{2}$ in the expression \eqref{eq_total_CS} play the role of filters for such values, this is a consequence of the approximation that has been used, a tree-level approximation.
One can naively find a singularity in $\mathbb{I}_{2}$ present in the term proportional to $\delta\left(\sqrt{\chi}\right)$ or even that $\delta\left(3\chi-1\right)$ leads to a imaginary contribution to the cross section. 
However, these observations are simply apparent, because when we analyze the full expression  for the cross-section \eqref{eq_total_CS} a natural condition arises upon these delta functions arise.
Once the cross-section represents a measurable physical quantity it must be real.
Therefore, applying this condition on \eqref{eq_total_CS}, we realize that the terms $\mathbb{I}_{1}$ and $\mathbb{I}_{2}$ must be real.
This implies a strong constraint upon the parameter $\chi$, i.e. $\chi>1$ (i.e. $\omega^2>m^2$).
Imposing this condition on $\mathbb{I}_{2}$, we realize that both delta functions $\delta\left(\sqrt{\chi}\right)$ and $\delta\left(3\chi-1\right)$ become zero and the cross-section depends only on the term $\mathbb{I}_{1}$.
Hence, the real and finite cross-section at finite temperature becomes
\begin{equation} 
\sigma_{\rm BW}	 	=\frac{e^{4}}{32\pi\omega^{2}}F^{2}\left(\beta\right)\mathbb{I}_{1}.\label{finite_CS}
\end{equation}

It is worth stress that the expression for the Breit-Wheeler process cross section \eqref{finite_CS} is valid for any value of $\beta$.
However, this thermal dependence is rather complicated in the factor $F^{2}\left(\beta\right)$.
Thus, one can consider, by means of elucidation, the high temperature limit, where one can get a better grasp of the thermal behavior of the cross section \eqref{finite_CS}.
Hence, at the high temperature regime, we obtain that
 \begin{align} \label{eq_high}
\sigma_{\rm BW}^{\rm high} \simeq\frac{e^{4}}{8\pi\omega^{2}}\mathbb{I}_{1}e^{-\frac{3\beta\omega}{4}}.
\end{align}
This expression is accordance with our expectation, since it is well known that at the high-temperature limit the Fermi-Dirac and Bose-Einstein distributions reproduce the Maxwell-Boltzmann statistics: $\frac{1}{e^{\beta\omega}\mp1} \simeq e^{-\beta\omega}$.

%Furthermore, it is important to note that the expression \eqref{eq_high} at finite temperature brings a different interpretation to the cross-section.
%Although the presence of the delta function in applications at finite temperature is common, here a new result can be discussed.
%The delta functions in the thermal expression $\mathbb{I}_{2}$  introduce a filter for such values, that is, the deltas specify certain values in the relation between momentum and energy. Then, this implies that the term $\mathbb{I}_{2}$ at finite temperature is an off-shell term.

\section{Conclusion}
\label{sec4}

In this paper we have studied thermal contributions to the Breit-Wheeler process within the thermofield dynamics approach.
Due to phenomenological interest (QED in intense electromagnetic fields) the Breit-Wheeler process, which consists of a pair production through the collision of two photons, has been considered in many contexts.
However, its description in a thermal bath is somewhat controversial, since many papers (results) disagree with each other.
Hence, we approach this topic within the thermofield dynamics formalism, which is very robust in order to shed some new light and clarity (some of) the controversies.

We started our analysis by reviewing the main aspects of the thermofield dynamics applied to the evaluation of the (thermal) differential cross section.
With this purpose, we discussed the second-order thermal transition amplitude related with the process $\gamma+\gamma \to e^- + e^+$, where the interacting part of the process is described by the usual QED coupling $e\Bar{\psi}(x)\gamma^\mu\psi(x)A_\mu(x) $ (and its dual).
One important point in the calculation of the transition amplitude is the use of the Bogoliubov transformation between the $T=0$ and thermal creation/annihilation operators, this is necessary because the in/out states are thermal states (operators), while the operators in the interacting Lagrangian are not thermal.
However, this is not the only thermal contribution to the scattering matrix, but the fermionic propagator also has a finite temperature part.

After establishing the main operations of the thermofield dynamics formalism, we considered the center of mass frame to compute the squared transition amplitude.
With this step we finally obtained the desired expression for the differential cross section Breit-Wheeler process at finite temperature.
In the zero temperature limit ($\beta\rightarrow \infty$) our results reproduced the known literature \cite{Bottcher:1997kf,torgrimsson:2018,Ribeyre}.
Remarkably, the thermal dependence of the differential cross section \eqref{eq181} had a simple form which allowed to compute the angular integration in order to obtain the Breit-Wheeler process cross section valid for any value of $\beta$.
Moreover, in order to have a better grasp about the thermal behavior of the cross section, we considered the high temperature regime, which led to a thermal dependence in terms of the Maxwell-Boltzmann distribution as we would expect.
%In addition, the thermal contribution leads to an off-shell term that selects some specific energy value. This is a new result at finite temperature that should be better understood in future studies.

\section*{Acknowledgments}

The authors thank Professor Ademir E. Santana for his interest and comments in the work.
This work by A. F. S. is partially supported by National Council for Scientific and Technological Develo\-pment - CNPq project No. 313400/2020-2. 
R.~B. acknowledges partial support from Conselho
Nacional de Desenvolvimento Cient\'ifico e Tecnol\'ogico (CNPq Project No. 306769/2022-0).

\global\long\def\link#1#2{\href{http://eudml.org/#1}{#2}}
 \global\long\def\doi#1#2{\href{http://dx.doi.org/#1}{#2}}
 \global\long\def\arXiv#1#2{\href{http://arxiv.org/abs/#1}{arXiv:#1 [#2]}}
 \global\long\def\arXivOld#1{\href{http://arxiv.org/abs/#1}{arXiv:#1}}

\end{document}